# LABORATORY ASTROPHYSICS WHITE PAPER

## (BASED ON THE 2010 NASA LABORATORY ASTROPHYSICS WORKSHOP IN GATLINBURG, TENNESSEE, 25-28 OCTOBER 2010)

Report prepared by the Scientific Organizing Committee (SOC):

Daniel Wolf Savin, *Columbia University*, SOC Chair
Lou Allamandola, *NASA Ames Research Center*
Steve Federman, *University of Toledo*
Paul Goldsmith, *NASA Jet Propulsion Laboratory*
Caroline Kilbourne, *NASA Goddard Space Flight Center*
Karin Öberg, *Harvard-Smithsonian Center for Astrophysics*
David Schultz, *Oak Ridge National Laboratory,* LOC Chair
Susanna Widicus Weaver, *Emory University*

Additional contributions on plasma laboratory astrophysics by:

Hantao Ji, *Princeton Plasma Physics Laboratory*
Bruce Remington, *Lawrence Livermore National Laboratory*

Reviewers:
Nancy Brickhouse, *Harvard-Smithsonian Center for Astrophysics*
Farid Salama, *NASA Ames Research Center*
Lucy Ziurys, *University of Arizona*

*Executive Summary*

Our understanding of the cosmos rests on knowledge of the underlying physical processes that generate the spectra, composition, and structure of the observed astronomical objects. Laboratory astrophysics is the Rosetta Stone that enables us to translate these observations into knowledge. Astrophysical discoveries such as the accelerating expansion of the universe, the atmospheric composition of exoplanets, and the organic nature of interstellar chemistry rely upon advances in laboratory astrophysics. This importance of laboratory astrophysics was noted by the National Research Council 2010 Decadal Survey of Astronomy and Astrophysics (Astro2010), which called out laboratory astrophysics as one of the NASA's core research programs "fundamental to mission development and essential for scientific progress".

The purpose of the 2010 NASA Laboratory Astrophysics Workshop (LAW) was, as given in the Charter from NASA, "to provide a forum within which the scientific community can review the current state of knowledge in the field of Laboratory Astrophysics, assess the critical data needs of NASA's current and future Space Astrophysics missions, and identify the challenges and opportunities facing the field as we begin a new decade". LAW 2010 was the fourth in a roughly quadrennial series of such workshops sponsored by the Astrophysics Division of the NASA Science Mission Directorate.

The findings of the 2010 LAW are that

- Astrophysical discovery continues to be propelled forward by advances in laboratory astrophysics leading, for example, to new insights into Type Ia supernovae, the role of water and complex organic molecules in interstellar chemistry, exo-planet formation, and accretion disk physics.

- Maximizing the scientific return of NASA's Space Astrophysics missions requires a vibrant laboratory astrophysics program to generate the needed data and ongoing support for databases to archive the data generated.

- Astro2010 recommended growing the Astrophysical Research and Analysis (APRA) program for laboratory astrophysics by $2 million over the baseline funding level.

- Astro2010 recommended that missions requiring significant amounts of new laboratory astrophysics data to reach their science goals should include within their program budgets adequate funding for the necessary experimental and theoretical investigations.

- NASA SMD Astrophysics Division support for laboratory astrophysics has fallen sharply. Since 2005, the number of awards has dropped by a factor of two. Funding has decreased by 32% since the 2006 LAW. Declining support is a prime candidate to explain the corresponding decrease in the number of submitted proposals.

- In order to maintain the basic laboratory astrophysics support required by current NASA Space Astrophysics missions, will require a restoration of funding levels to their previous levels. Additional funding to supplement this basic level of support is needed to enable the laboratory astrophysics community to develop the new techniques required to provide critical support for the next generation of NASA Space Astrophysics missions.

The required laboratory astrophysics science for current and future NASA Space Astrophysics missions span from the sub-mm to the X-ray and include every bandpass in between. The needed information includes data for spectra and structure; electron driven collisions; heavy particle collisions; photon driven processes; chemical reactions in the gas phase, on grain surfaces, and in ices; formation and destruction mechanisms for molecules, dust, and ices; and radiatively driven, magnetically driven, and kinetically driven plasma processes  Addressing a number of these needs will require the development of new technologies and instrumentation and, in some cases, interagency collaboration involving various combinations of NASA, NSF, and the Departments of Commerce, Defense, and Energy.

In order to realize the recommended scientific goals of the 2010 Decadal Survey, the 2010 LAW found that there were a number of possible concrete actions which NASA could take.  These include:

- **Restoring APRA funding to the baseline funding level of FY 2006.**

- **Implementing the Astro2010 recommendations to grow the APRA program support and to provide mission support for laboratory astrophysics.**

- **Establishing a series of new initiatives in order to revitalize, grow and ensure the future of laboratory astrophysics:**

- **Developing an appropriate mechanism to ensure the long-term viability of laboratory astrophysics databases.**

- **Continuing to sponsor this quadrennial series of Laboratory Astrophysics Workshops, the next meeting of which should take place in the 2014-2015 period.**

*1. Preface*

"Laboratory astrophysics and complementary theoretical calculations are the foundations of astronomical and planetary research and will remain so for many generations to come. From the level of scientific conception to that of the scientific return, it is our understanding of the underlying processes that allows us to address fundamental questions regarding the origins and evolution of galaxies, stars, planetary systems, and life in the cosmos. In this regard, laboratory astrophysics is much like detector and instrument development at [NASA]; these efforts are necessary for the astronomical research being funded by [NASA]." This opening paragraph from the White Paper generated by the 2006 NASA Laboratory Astrophysics Workshop (LAW) remains as valid now as it was then.

The 2010 NASA LAW was held in Gatlinburg, Tennessee, from 26-28 October 2010. It was sponsored by the Astrophysics Division of the NASA Science Mission Directorate (SMD). The Charter from NASA is given in Appendix A and the format of the workshop is described in Appendix B. As was done in the past, laboratory astrophysics is taken to encompass both laboratory and theoretical work and no distinction is made between the two unless necessary.

The Workshop was timed to take place shortly after the National Research Council 2010 Decadal Survey on Astronomy and Astrophysics (Astro2010) issued their report. In this way, LAW 2010 would be able to respond to and amplify upon the Astro2010 recommendations.

The Astro2010 Decadal Survey stressed the critical importance of laboratory astrophysics for advancements in astronomy and in the science return from space missions. The survey issued strong recommendations to augment the funding of the APRA program by $2M/year and that missions which "require significant amounts of new laboratory data to reach their science goals should include within their program budgets adequate funding for the necessary experimental and theoretical investigations."

Throughout this White Paper, as requested in the Charter, we "outline specific opportunities and threats facing NASA's Laboratory Astrophysics Program, and articulate concrete actions by which the Agency can capitalize on the opportunities and mitigate the challenges."

*2. General Findings*

- Astrophysical discovery continues to be propelled forward by advances in laboratory astrophysics.

- Maximizing the scientific return of NASA's Space Astrophysics missions requires a vibrant laboratory astrophysics program to generate the needed data and ongoing support for databases to archive the data generated.

- Astro2010 recommended growing the Astrophysical Research and Analysis (APRA) program for laboratory astrophysics by $2 million over the baseline funding level.

- Astro2010 recommended that missions requiring significant amounts of new laboratory astrophysics data to reach their science goals should include within their program budgets adequate funding for the necessary experimental and theoretical investigations.

- NASA SMD Astrophysics Division support for laboratory astrophysics has fallen sharply. Since 2005, the number of awards has dropped by a factor of two. Funding has decreased by 32% since the 2006 LAW. Declining support is a prime candidate to explain the corresponding decrease in the number of submitted proposals.

- In order to maintain the basic laboratory astrophysics support required by current NASA Space Astrophysics missions, will require a restoration of funding levels to their previous levels. Additional funding to supplement this basic level of support is needed to enable the laboratory astrophysics community to develop the new techniques required to provide critical support for the next generation of NASA Space Astrophysics missions.

*3. Recent Successes*

Astrophysical discovery continues to be driven in part by experimental and theoretical work in laboratory astrophysics. For example, atomic astrophysics data have been critical for studying supernovae over a wide wavelength range. They have also played a key role in observations of O stars, and are having an impact on wavelength standards used for exoplanet searches employing wobble-induced Doppler shifts. Molecular astrophysics has been essential for interpreting data from the Herschel Space Observatory on a number of key species, notably water. Molecular data have underlain impressive results on recently detected fullerene molecules, and on the spectroscopy of hot-Jupiter exoplanets. Dust and ices play many important roles in the interstellar medium, and study of them requires data on broadband opacities and spectral features, as well as laboratory data on different types of surfaces that impact chemistry in dense regions. Plasma astrophysics looks at diverse regions of high energy density, studies of which require a broad range of laboratory results. These address problems ranging from planet formation to accretion disks. Appendix C elaborates on these examples of the richness of astrophysical discovery arising from laboratory astrophysics.

*4. Current and Future Needs*

Current and future space astrophysics missions span from the sub-millimeter to the X-ray and include every bandpass in between. Laboratory astrophysics capabilities need to cover all of these bandpasses. As highlighted below, other common needs for laboratory astrophysics are accurate line intensities, useful databases, and high-performance computing capabilities (to study complex systems and to provide benchmarks). Addressing a number of the needs listed below will require the development of new technologies and instrumentation. Some will also require interagency collaboration involving various combinations of NASA, NSF, and the Departments of Commerce, Defense, and Energy. Examples of areas for which such cooperation is needed include databases (as is discussed in Section 5.2) and plasma astrophysics (which often require experimental facilities only available at DOE laboratories).

The rest of this section is largely in bullet format and not in any order of priority. Space limitations prevent us from providing more than a general listing of the required data. This section is also likely incomplete because of our inability to fully anticipate the future. Proposers seeking NASA Astrophysics funding for laboratory astrophysics can use the points below as a guide, but they will need to make a clear connection between the astrophysical science problem(s), the relevant NASA Space Astrophysics mission(s), and how the specific proposed laboratory astrophysics research will help to achieve the desired astrophysical understanding.

*4.1 Atomic Astrophysics*

Data are needed for all cosmically important elements observed by NASA space astrophysics missions. This includes not only elements with atomic number $A<30$, but also neutron capture elements with $A>30$. Of particular importance are the iron group elements, which are ubiquitous, and low charge states of *r*- and *s*-process elements which are important for understanding nucleosynthesis during the end points of stellar evolution. All these data are often needed for ground states and low-lying metastable levels of both neutrals and ions. Reliable error estimates are needed for all data generated.

*4.1.1 Spectra and Structure*
- Obtain better wavelengths, line identifications, oscillator strengths, hyperfine structure, and isotope shifts for the first 3 spectra of iron group elements and beyond.
- Improve wavelengths and oscillator strengths for fine structure lines above 1 μm.
- Measure X-ray wavelengths accurate to better than 1 mÅ for Fe and Ni L- and M-shell spectra, including inner-shell transitions.
- Determine wavelengths for satellite lines generated through dielectronic recombination.
- Improve accuracy of line diagnostics in the X-ray region.

*4.1.2 Electron-Impact Collisions*
- Improve electron impaction excitation data for Fe and Ni L-shell ions, H- and He-like ions for high excitation to high Rydberg levels, temperature and density sensitive lines, low charge states of Fe group elements, and forbidden transitions.
- Improve electron impact ionization data for valence and inner-shell electrons, especially benchmark laboratory measurements using ion beams either free of metastable contamination or with well characterized initial populations.
- Generate reliable dielectronic recombination data for low temperature plasmas and in high temperature plasmas for satellite line strengths and wavelengths.
- Develop modern data for Auger yields resulting from innershell ionization or excitation.
- Continually update ionization equilibrium calculations including sensitivity studies of how uncertainties in the ionization and recombination data propagate through to the predicted fractional abundances.

*4.1.3 Heavy Particle Collisions*
- Generate state-of-the-art charge exchange data for ions with neutrals such as H, He, $H_2$, $CO_2$, $H_2O$, etc.
- Improved fine structure excitation cross sections due to ion collisions with H, He, and $H_2$.
- Improve electronic excitations data for neutral-neutral collisions with H, He, and $H_2$.

- Calculate modern proton impact excitation and ionization data.
- Determine reaction rates between H, N, O, and S for PAHs containing between 50 and 200 carbon atoms in the neutral, cation, and anion states.
- Determine $H_2$ formation rates on PAHs containing between 50 and 200 carbon atoms as a function of PAH size and charge.

*4.1.4 Photon Driven Processes*
- Obtain better photoionization data of innershell K and L electrons including details around the K-shell threshold.
- Generate reliable Auger and fluorescence yields due to innershell photo-excitation or ionization and provide data as a function of both initial and final state.
- Perform experiments at synchrotron light sources and free electron lasers on highly charged ions to benchmark theory.
- Obtain probability distributions for the number of electrons emitted following the creation and then relaxation of an innershell hole.

*4.2 Molecular Astrophysics*

*4.2.1 Instrument and Technology Development*
- Develop broadband measurement techniques to enable rapid data acquisition over large ranges in frequency.
- Develop new techniques to address the challenge of generating, processing, monitoring and measuring in the laboratory the spectra and the structure of large complex molecules under conditions analogous to space conditions.
- Increase sensitivity (especially in the far-IR) to enable characterization of unstable species.
- Increase wavelength coverage to match NASA mission capabilities.
- Develop new methods for reaction kinetics and dynamics measurements.

*4.2.2 Spectral Complexity*
- Expand lab efforts beyond the current focus on interstellar "weeds" (commonly observed molecules with many lines in a frequency interval) to also include interstellar "flowers" (species of particular importance).
- Increase spectral coverage for all molecules, especially above 2 THz through mid-IR.
- Develop better theoretical models that couple rotation and vibration.
- Generate more complete sets of data on hot band lines.

*4.2.3 Molecular Complexity*
- Bridge gap between molecular astrophysics (small molecules with up to ~tens of atoms) and solid-state astrophysics (large molecules such as PAHs with >>100 atoms).
- Develop new approaches for interpreting spectra and excitation of complex molecules.
- Identify wavelength ranges for measurements of signatures of complex molecules.
- Extend chemical models to include large species.
- Determine electron affinities and electron-ejection energies for PAHs containing between 50 and 200 carbon atoms.

*4.2.4 Spectral and Kinetic Databases*
- Streamline.
- Maintain.
- Critically review.
- Extend.

*4.2.5 Science Interpretation*
- Conduct kinetics measurements for key reactions.
- Develop new methods and extend existing methods to study reaction dynamics.
- Measure collisional excitation rates.
- Quantify line intensities.
- Extend chemical models.

*4.2.6 Computation and Theory*
- Complete work on small molecules.
- Extend methods to large molecules.
- Determine reaction rates.
- Develop high performance computing contributions to spectroscopy and interpretation.
- Provide laboratory and computational benchmarks.

4.3 Dust and Ices Astrophysics

4.3.1 Ice, Dust, and PAH Identification at IR Wavelengths
- Prepare and characterize, spectroscopically and structurally, standard (terrestrial) and astronomically realistic silicatious, carbonaceous, ice (containing PAHs), and dust analogs to understand the differences between terrestrial "standard" and extraterrestrial materials. Characterize these materials both before and after *in-situ* energetic processing with UV and high energy particles (simulating cosmic rays). Measure the UV, optical, mid- and far-IR spectra of these materials from 0.1 to 1000 µm and determine optical constants as a function of composition and temperature.
- Measure the UV, optical, mid-and far-IR spectra of PAHs from 0.1 to 1000 µm with a focus on PAHs containing between 60 and 200 carbon atoms. Spectra are needed for neutral, cation, and anion forms. Investigations into the effects of oxygen and nitrogen incorporation into PAHs are also important.
- Measure the UV, optical, mid- and far-IR spectra of homo- and hetero-geneous PAH clusters as a function of cluster size and charge.
- Establish the protocol and standards for the creation of an UV to sub-millimeter spectroscopic database of refractory dust materials, pure ices and ice mixtures, and develop the database.

4.3.2 Diagnostics of Surface Formation Pathways
- Measure reactions comprising the most abundant, reactive elements H, C, N, O, S, and P on different types of dust surfaces at different temperatures and determine isotopic fractionation, new radical and molecule formation and destruction routes, and their branching ratios.
- Determine the rotational temperatures of the various ice species as they are ejected into the gas phase from realistic ice/dust surfaces at different temperatures.

4.3.3 Ice Formation and Destruction
- Develop a comprehensive experimental program to measure the yields of various reaction pathways, including dissociation channels and desorption mechanisms, as a function of input energy.
- Create models of the chemical evolution during star and planet formation with a sufficient level of detail to interpret current observations.
- Measure ice and carbonaceous dust formation and destruction processes (including shock and sputtering pathways) under realistic protoplanetary disk conditions to determine the amount of volatiles and carbon generated.

*4.3.4 PAH and Amorphous Carbon Particle Formation and Destruction*
- Delineate PAH formation pathways under circumstellar conditions as a function of C/H ratios.
- Delineate carbonaceous particle formation routes from molecular precursors in circumstellar shells and determine the factors that determine their final H/C ratio and aromatic to aliphatic C ratio.
- Determine PAH destruction mechanisms and kinetics in various astronomical environments (HII regions, PNe, PDRs, etc.) as a function of molecular size.

*4.4 Plasma Astrophysics*

*4.4.1 Radiatively Driven Laboratory Experiments*
- Develop scaled, radiatively driven, nonlinear hydrodynamics experiments suitable to test astrophysical models of molecular cloud dynamics and star formation dynamics.
- Extend precision opacity measurements into stellar core conditions.
- Extend equation of state measurements to higher pressures and off-Hugoniot conditions more relevant to star and planet formation.
- Develop non-relativistic and relativistic collisionless shocks to test our fundamental understanding of shocks in the universe.
- Extend scaled supernova explosion dynamics into the turbulent regime in relevant geometries to test astrophysical modeling of nature's most powerful explosions.
- Develop experiments of shock and radiation processing of the environment (dust, ice, molecules) to improve our understanding of the formation of stars, planets, and life.

*4.4.2 Magnetically Driven Laboratory Experiments*
- Build next generation experiments to study reconnection in larger and more collisionless plasmas to be directly relevant to space and astrophysical plasmas, as well as the associated particle acceleration.
- Explore reconnection processes in high energy density (HED) conditions where plasma kinetic energy substantially exceeds magnetic energy.

- Investigate the effects of plasma processes and magnetic fields on the dynamics of dust particles, including charging, growth, breakup, and collective processes.
- Provide EUV spectroscopic data of metal inner-shell lines at high density and well-confined fusion plasmas.

*4.4.3 Kinetically Driven Laboratory Experiments*
- Develop experimental capabilities to access and understand nonlinear dynamos at larger magnetic Reynolds numbers in liquid metal and plasmas.
- Detect and study magnetorotational instability and the associated angular momentum transport scaling with respect to magnetic Reynolds number in liquid metal, gas, and plasmas.
- Extend experiments of flow driven and jet driven shocks to higher Mach number and larger size (Reynolds number) to test modeling of star formation dynamics.
- Develop experiments to test our understanding of the role of magnetic fields in instabilities, mixing, and evolution of turbulence.

*5. Specific Issues/Questions for Consideration*

*5.1 What steps should be taken to increase collaboration, cooperation, and communication within the NASA Laboratory Astrophysics Program?*

One suggestion at the workshop included the establishment of a database listing data needed by the space mission astrophysics community. One form that this could take would be a listserv or wiki to which people could post requests. Other suggestions included establishing a quarterly newsletter and/or using social networking tools such as facebook and twitter. The importance of venues where people could come together was noted.

To a certain extent, some of these suggestions are already in place. This is a direct outgrowth of the 2006 LAW which recommended that the American Astronomical Society (AAS) establish a Working Group on Laboratory Astrophysics (WGLA). The WGLA was established in May 2007 and now sponsors a yearly meeting on laboratory astrophysics which is held during the AAS summer meeting. To date the WGLA has sponsored 4 such meetings, the first on laboratory astrophysics as a whole, the second on dust and ices, the third on plasmas, and the upcoming one on nuclear and particles astrophysics. Their listserv currently has over 300 members. Building upon the successes of the WGLA would go a long way toward increasing collaboration, cooperation, and communication within the NASA Laboratory Astrophysics Program.

Databases and the required collaborations needed to maintain them also provide an opportunity to achieve this goal. So would large collaborative programs which are instrument heavy. However, each of these requires sufficient levels of funding to be viable. Another suggestion involved having close ties between laboratory astrophysics, observational programs, and mission user groups. This could be done, in part, though the creation of mission-specific laboratory astrophysics programs which would receive their support from the missions.

*5.2 How should laboratory astrophysics databases be organized to minimize redundancy and cost, validated to ensure data consistency and quality, and curated to guarantee easy, widespread access by the community of data users?*

The ultimate answer to this question will likely involve participation by the relevant domestic, foreign, and international agencies, departments, and organizations. To achieve this will likely require a NASA-sponsored workshop involving all relevant parties in order to determine how best to proceed.

Databases need to be maintained over periods of decades. Appropriate funding strategies are needed to ensure their survival. Mission support can help in part, but the databases clearly have lifetimes far longer than any NASA astrophysics space mission and their applications span multiple missions. Funding from the Astrophysics Data Curation and Archiving (ADCAR) program would be a more stable source of long term support. The databases also need to stay current and of relevance to the astrophysics users. Databases should strive to be compatible with one another and with the principles of the Virtual Observatory (e.g., easy public access with sufficient metadata to enhance portability, track version number, and proper data source attribution). While the data do need to be critically evaluated before inclusion in the database, often unevaluated data are better than no data and the databases should be structured to accommodate such data. Previous versions of databases also need to be available so that researchers can readily determine how astrophysical interpretations have changed between the old and new data.

Laboratory astrophysics databases which are commonly used by astrophysicists are maintained in the U.S. by NASA, the Department of Commerce, the Department of Energy, and the Department of Defense. A number of foreign and international organizations also maintain relevant databases. All together, the various databases in this country and abroad overlap to varying degrees with the NASA space astrophysics needs, but the overlap is rarely perfect. Additionally there is no standard format used across databases for the archived information. Organizing all these databases so as to minimize redundancy and cost, validate the data to ensure consistency and quality, and curate the data to guarantee easy, widespread access by the community of data users represents a grand challenge which is going to require a significant amount of continued study. A model for consideration is that established by the NASA ADCAR program or by the Department of Energy for nuclear data (U.S. Nuclear Data Program – USNDP), involving national laboratories, universities, and NIST. However, the budget for USNDP exceeds the current APRA budget by nearly a factor of two

*5.3 How might a new initiative in supporting laboratory astrophysics research (e.g. group/team awards, centers of excellence, NASA Laboratory Astrophysics Institute) be structured and implemented so as to augment the current program?*

The emphasis here is on how such an initiative could *augment* the current laboratory astrophysics research. The LAW community felt very strongly that such an initiative should not come at the expense of the APRA program which, through its support of individual Principal Investigators (PIs), helps to maintain the core competency of the laboratory astrophysics community.

That said, the community recognized that there are some scientific questions which are too big for a single PI to solve and which require a multi-expertise approach. To that end, NASA could establish a program of group/team awards in laboratory astrophysics to address well-defined grand astrophysical challenges. The questions posed by Astro2010 provide examples of such challenges. This program should be of limited term. It was felt that five years was a reasonable length. Proposals to the program would have to lay out a specific set of goals to be accomplished in that time. Any renewals would have to compete in an open call for proposals to ensure that the supported research remains relevant to NASA space astrophysics missions. Concern was also expressed that such a program would have to be structured to prevent there being too many group members on a proposal, thereby resulting in none of the research groups getting sufficient funding to carry out the proposed research.

*5.4 What is the role of NASA's field centers and scientists in pursuing the Agency's critical Laboratory Astrophysics data needs?*

While there was not a lot of discussion of this question at the meeting, some points were raised on the many possible roles for NASA's field centers and scientists in pursuing the critically needed laboratory astrophysics data. These include identifying and articulating NASA's needs for various missions, generating the needed data, making their unique research facilities available to the broader community for collaboration, and technology transfer of state-of-the-art instrumentation (particularly detectors) to non-NASA scientists. Although, to varying degrees, some of this is already largely occurring, there is clearly room for expansion.

*5.5 What steps can NASA take to attract new talent (e.g. graduate student/postdoctoral fellowships, targeted proposal opportunities for junior faculty)?*

Bringing new talent to laboratory astrophysics will require providing funding opportunities at all career levels. Establishing both graduate student and postdoctoral research fellowships in the field will help to direct the recipients into careers in laboratory astrophysics. The transition from postdoctoral researcher to either a research or faculty track position would be aided by targeted proposal opportunities in laboratory astrophysics for outstanding young investigators who have completed their postdoc and are less than 10 years from having received their Ph.D. or equivalent. A robust APRA program coupled with a robust technology development and instrumentation program that also funds the required personnel will then enable these researchers to build and maintain their research programs and thereby continue their careers in laboratory astrophysics. Mission supported programs in laboratory astrophysics would augment this. These last three programs will also provide opportunities for researchers and faculty in other related areas to redirect their research into laboratory astrophysics and focus on issues important to NASA's Space Astrophysics missions. In short, if sufficient funding is provided for the field, that will attract new talent at all career levels.

*6. LAW 2010 Conclusions*

In order to realize the recommended scientific goals of the 2010 Decadal Survey, the 2010 LAW found that there were a number of possible concrete actions which NASA could take. These include:

**1) Restoring APRA funding to the baseline funding level of FY 2006.** The core competency of laboratory astrophysics needed to maximize the scientific return from NASA space astrophysics missions has eroded significantly over the past 4 years. If the current funding trend is not reversed, the field will die within the decade. Returning the baseline funding level to that of FY 2006 will reverse the trend and repair the damage.

**2) Implementing the Astro2010 recommendations to grow the APRA program support and to provide mission support for laboratory astrophysics.** These recommendations are critical to maintaining a core competency in laboratory astrophysics combined with enabling the field to respond on a "rapid" time scale to specific NASA space mission needs. The Astro2010 recommended $2M growth of the APRA program should be in addition to the restoration of the program to its baseline funding level of FY 2006. The Astro2010 recommendations echo two of the findings from the 2006 LAW White Paper. The WGLA, another product of the 2006 LAW, emphasized these points in the White Papers they submitted to each of the Astro2010 Science Frontier Panels (SFPs) as well as to the Astro2010 Infrastructure Working Group on Facilities, Funding, and Programs. That these 2006 LAW findings made it into the final Astro2010 recommendation testifies to the importance of their implementation.

**3) Establishing a series of new initiatives in order to revitalize, grow, and ensure the future of laboratory astrophysics:**
- Graduate student and postdoctoral research fellowships in laboratory astrophysics are needed to train the next generation in the field. These fellowships should be tenable at any U.S. host institution of the fellows' choice.
- A technology development and instrumentation program that also supports the required personnel is needed to enable vital research which cannot be performed with current facilities. This program should be open to both experimentalists and theorists.
- Targeted proposal opportunities in laboratory astrophysics are needed for outstanding young investigators who have completed their postdoc but are less than 10 years from having received their Ph.D. or equivalent. Such a program should be open to young investigators in either a research or faculty track at their host institution.
- As part of Phase B plans for future missions, laboratory data requirements to achieve the scientific goals described in the mission proposal need to be addressed, with sufficient funding in the budget to acquire the necessary data.
- A program of group/team awards in laboratory astrophysics is needed to address grand astrophysical challenges. This program should be of limited term and proposals would have to lay out specific goals to be accomplished in that time. Renewals would have to compete in an open call for proposals to ensure that the supported research remains relevant to NASA Space Astrophysics missions.
- None of these new initiatives should come at the expense of the APRA program which supports the core competency of the field.

**4) Developing an appropriate mechanism to ensure the long-term viability of laboratory astrophysics databases.** Databases are vital archives for the vast quantities of laboratory astrophysics data which have been and are being generated. If these data are not properly archived, they will be lost to future generations as the producers of the data retire and the corresponding facilities are decommissioned. Regenerating the needed data often requires years of development at a cost vastly exceeding that of the original work. Laboratory astrophysics databases, which are commonly used by astrophysicists, are maintained by a variety of domestic, foreign, and international organizations, agencies, and departments, the needs of which overlap to varying degrees with the NASA space astrophysics needs. The ultimate solution to ensuring the long-term viability of the laboratory astrophysics data is likely to involve cooperation between the various domestic, foreign, and international agencies, departments, and organizations who currently maintain the relevant databases. Possible models include NASA's ADCAR program or the USNDP. To address this issue will likely require a NASA-sponsored workshop involving all relevant parties in order to determine how best to proceed.

**5) Continuing to sponsor this quadrennial series of Laboratory Astrophysics Workshops, the next meeting of which should take place around 2014-2015.** Astro2010 laid out three grand science objectives for the coming decade and emphasized the importance of laboratory astrophysics to enable the successful accomplishment of these goals. The various SFPs each laid out four questions for the coming decade and one discovery area. Four of the five SFPs emphasized the importance of laboratory astrophysics in pursuing these goals. As NASA Space Astrophysics missions work to achieve the objectives laid out by Astro2010 and the associated SFPs, it is clear that laboratory astrophysics will play a critical role. For this reason, it is vital that regular LAWs be held throughout the coming decade to monitor the health of the field and to ensure NASA's Space Astrophysics missions can meet the objectives laid out in the Astro2010 Decadal Survey.

Appendix A: Charter for the 2010 NASA LAW

**Purpose of workshop:** The purpose of the Laboratory Astrophysics Workshop (LAW) 2010 is to provide a forum within which the scientific community can review the current state of knowledge in the field of Laboratory Astrophysics, assess the critical data needs of NASA's current and future space astrophysics missions, and identify the challenges and opportunities facing the field as we begin a new decade. LAW 2010 is sponsored by the Astrophysics Division of NASA's Science Mission Directorate (SMD).

**Target audience:** Laboratory Astrophysicists and Astrochemists (experimentalists, theorists, and modelers), Astronomers and Astrophysicists (observers, theorists, and modelers), Space Mission Scientists, Instrument Developers and other interested researchers.

**Specific Goals:** LAW 2010 will:
1. Review the current state-of-the art in laboratory astrophysics;
2. Review the recommendations of previous LAWs and assess progress toward meeting those recommendations.
3. Identify the critical data needs of NASA's current and future planned space astrophysics missions, and assess the degree to which NASA-supported research efforts currently address those data needs;
4. Assess the strengths, weaknesses, opportunities, and threats facing NASA's Laboratory Astrophysics program in the context of the Astro2010 Decadal Survey report;
5. Formulate a White Paper summarizing the key findings from the workshop for submission to the NAC Astrophysics Subcommittee and the Astrophysics Division (draft to be submitted by 25 Jan 2011; final report to be submitted by 25 Feb 2011);
6. Generate a volume of science proceedings from the workshop that will serve as a reference to NASA and the community, and distribute that volume through the NASA Astrophysics Data System (ADS).

**Overview:** LAW 2010 is the fourth in a series of NASA-sponsored Laboratory Astrophysics Workshops. Held approximately quadrennially, previous LAWs were held in 1998 (Harvard-Smithsonian Center for Astrophysics), 2002 (NASA Ames Research Center), 2006 (U. Nevada, Las Vegas). The strength of these workshops lies in bringing together producers and users of laboratory astrophysics data so that they can understand each other's needs and limitations in the context of NASA's mission needs. The workshops also serve to increase collaboration and cross fertilization of ideas thereby ensuring that the priorities of NASA's Laboratory Astrophysics Program are aligned with the critical data needs of NASA's space astrophysics missions, and that the products of NASA-sponsored Laboratory Astrophysics research feeds back to the user community in a timely way.

The single most important and valuable deliverable from LAW 2010 will be the White Paper summarizing the proceedings and outcomes of the workshop. That White Paper should provide detailed findings on the critical laboratory astrophysics data that are required to maximize the scientific return on NASA's past, current, and future planned space astrophysics missions. The White paper should also outline specific opportunities and threats facing NASA's Laboratory

Astrophysics Program, and articulate concrete actions by which the Agency can capitalize on the opportunities and mitigate the challenges.

Another important item in the White Paper should be a tabulation of recent significant astronomical results where the input from laboratory astrophysics was of critical importance (although the laboratory astrophysics contribution may not have received the credit due it, as is so often the case). The current funding environment, including the very difficult years ahead, will require a certain amount of salesmanship.

Lastly, with an eye on the future, a discussion should also be given as to what can be done in order to help foster the creation of new faculty positions and the education and production of future generations of laboratory astrophysics scientists.

**Specific Issues/Questions for Consideration:**
- What steps could be taken to increase collaboration, cooperation, and communication within the NASA Laboratory Astrophysics Program?
- How should laboratory astrophysics databases be organized to minimize redundancy and cost, validated to ensure data consistency and quality, and curated to guarantee easy, widespread access by the community of data users?
- How might a new initiative in supporting laboratory astrophysics research (e.g. group/team awards, centers of excellence, NASA Laboratory Astrophysics Institute) be structured and implemented so as to augment the current program?
- What is the role of NASA's field centers and scientists in pursuing the Agency's critical Laboratory Astrophysics data needs?
- What steps can NASA take to attract new talent (e.g. graduate student/post doctoral fellowships, targeted proposal opportunities for junior faculty).

**Agenda:** The workshop will feature invited talks by members of the scientific community representing data users. The invited speakers will provide a broad overview of the needs in the field. The workshop will also feature shorter talks and posters by data producers (NASA Laboratory Astrophysics Program grantees and others). Breakout sessions chaired by the Scientific Organizing Committee members will be held to promote discussion of focused issues of importance to meeting the critical data needs of NASA space astrophysics missions.

Appendix B: Workshop Format

The Workshop began with a report from a Survey Committee member, followed by presentations given by one member from each of the five Astro2010 Science Frontier Panels (SFPs). In addition there were four invited review talks on current and future NASA Space Astrophysics Missions as well as 13 on various areas in laboratory astrophysics. These 23 speakers represented both users and producers of laboratory astrophysics data. Sessions for contributed posters were held on the first two days of the Workshop, highlighting exciting new developments in laboratory astrophysics. The number of laboratory astrophysics grants awarded by the Astrophysics Division over the 5 funding cycles preceding LAW 2010 is roughly 47% atoms, 38% molecules, 12% solids, and 4% plasma. The number of the poster presentations closely reflected this distribution. On the third day, breakout sessions in each of these four areas provided an opportunity for users and producers to discuss current and future laboratory astrophysics needs for NASA space astrophysics missions. The meeting concluded with reports given by each of the breakout groups followed by a plenary discussion of the findings.

Appendix C: Recent Successes

Astrophysical discovery continues to be driven in part by experimental and theoretical work in laboratory astrophysics. Here we present selected examples of significant astrophysical advances arising from recent laboratory astrophysics studies. These examples demonstrate the richness of astrophysical discovery arising from laboratory astrophysics but are not meant to be exhaustive; space limitations prevent us from being complete.

*C.1 Atomic Astrophysics*

Type Ia supernovae (SNe) are used as standard candles to study dark energy and the expansion of the universe. Chandra and XMM-Newton X-ray studies of young supernova remnants (SNRs) have deepened our understanding of these standard candles. X-ray spectra of young SNRs in the Milky Way and the Magellanic Clouds offer the most detailed view of Type Ia SN ejecta available at any wavelength and provide invaluable constraints on the physics of these explosions and the identity of their progenitor systems. Utilizing public domain atomic data, it is now possible to model the X-ray emission and distinguish SNRs resulting from bright and dim Type Ia SNe. This has been validated by the detection and spectroscopy of SN light echoes for the Tycho SNR (Badenes et al. 2006, ApJ, 645, 1373; Krause et al. 2008, Nature, 456, 617) and SNR 0509-67.5 in the LMC (Badenes et al. 2008, ApJ, 680, 1149; Rest et al. 2008, ApJ, 680, 1137). Key advantages of these X-ray studies of nearby SNRs over optical studies of extragalactic SNe are that the SNRs are close enough to examine the circumstellar medium sculpted by the progenitor systems (e.g., the Kepler SNR, Reynolds et al. 2007, ApJ, 668, L135) and also to study the resolved stellar populations associated with them (Badenes et al. 2009, ApJ, 700, 727). Recent X-ray observations have also discovered emission from Mn and Cr in young Type Ia SNRs. This can be used to measure the metallicity of the progenitor system (Badenes et al. 2008, ApJ, 680, L33), one of the key variables that might affect the cosmological use of Type Ia SNe and which cannot be determined for extragalactic SNe.

Advances in our understanding of the elemental evolution of the cosmos has come about from spectroscopic observations carried out using HST, Chandra, and XMM-Newton coupled with new laboratory astrophysics data. For example, a clearer distinction between the main *r*-process and the "weak" *r*-process has resulted from new transition probabilities for many neutron capture elements. This finding rests on detailed abundance studies of the neutron capture elements in old metal-poor halo stars carried out using both HST-STIS and ground-based observations including Keck I HIRES (e.g., Sneden et al. 2009, ApJS, 182, 80) coupled with the measurement of new transition probabilities for many neutron capture elements (e.g., Lawler et al. 2009, ApJS, 182, 51). The *r*-process neutron capture elemental abundance pattern which has emerged from this effort constrains nucleosynthesis modeling and provides a clearer distinction between the main *r*-process and the weak *r*-process (Kratz et al. 2007, ApJ, 662, 39).

Another example involves X-ray observations of O stars. Their powerful radiatively driven winds are important sources of chemical enrichment in the universe. Recent analyses of UV P Cygni profiles and X-ray emission line profiles have been used to determine mass loss rates (Fullerton et al. 2006, ApJ, 637, 1025; Cohen et al. 2010, MNRAS, 405, 2391). These studies used state-of-the-art wavelengths (accurate to a few mÅ), and a relatively complete database of important X-ray emission lines, together with data on relative line strengths in coronal plasmas, in order to accurately account for blended complexes of Doppler broadened emission lines. These works found that the mass loss rate from O-stars is a factor of a few less than previously thought, a finding resulting from recent improvements in atomic data from laboratory and theoretical calculations. This changes our understanding of chemical enrichment of galaxies, especially during their early starburst phase.

Exoplanet discovery from planetary-induced stellar line Doppler shifts rests on precisely calibrated wavelength standards. These discoveries depend on the measurement of small changes in the wavelengths of the stellar lines due to the Doppler shift caused by the wobble of the star as the planet moves around it. A velocity change of 10 m s$^{-1}$ requires the measurement of Doppler shifts to 3 parts in $10^8$. To achieve this precision, an accurate wavelength reference is needed that is stable over decades of time (Marcy & Butler 1992, PASP, 104, 270). The HIRES spectrograph on the Keck telescope, used in the recent discovery of a planet in the habitable zone of a star (Vogt et al. 2010, ApJ, 723, 954), uses a carefully calibrated iodine reference cell placed in the beam from the telescope. The calibration of this cell is made in the laboratory using high-resolution Fourier transform spectroscopy. Similar discoveries using the HARPS spectrograph in Chile use thorium-argon hollow cathode lamps that have been calibrated in the laboratory.

*C.2 Molecular Astrophysics*

Water plays a key role in interstellar chemistry. It is the dominant molecular carrier of oxygen in interstellar clouds, and it is important in biological systems on Earth. Far-infrared observations using Herschel now open the possibility for direct observations of water and related molecules. First results from the Herschel Key Programs indicate the exciting possibilities in store for advancing our understanding of the water budget during star and planet formation. O-related chemistry has challenged interstellar cloud chemical models. Odin Satellite observations of water in dense clouds gave strict upper limits on the fractional abundance relative to $H_2$ (Klotz et al. 2008, A&A, 488, 559), which stood in stark contrast to model predictions (Lee et al. 1996, A&AS, 119, 111). The abundance of water in cold clouds and its formation and destruction remained a mystery until recent Herschel observations. They provided the first direct detection of water in starless cores, such as in L1544 (Caselli et al. 2010, A&A, 521, L29). Surprisingly, water was not detected in the DM Tau protoplanetary disk (Bergin et al. 2010, A&A, 521, L33). However, Herschel results reveal a rich water-related chemistry in other types of interstellar environments. Ionized water, $H_2O^+$, and the related ion $OH^+$ were detected in numerous sources ranging from massive star-forming cores (Schilke et al. 2010, A&A, 521, L11; Gupta et al. 2010, A&A, 521, L47 2010) to diffuse clouds (Neufeld et al. 2010, A&A, 521, L10). Far-IR observations of water using Herschel also enabled the first determination of the water production rate from a comet (Val-Borro et al. 2010, A&A, 521, L50). Likewise, the isotopic fractionation of water in massive star-forming regions was traced through Herschel observations (Comito et al. 2010, A&A, 521, L38). These advances in the understanding of the water budget during star and planet formation were based on laboratory studies of water and related species in the submillimeter range. The spectral catalog for water was extended to include transitions across the frequency range covered by the HIFI instrument on Herschel (Pickett et al. 2005, JMS, 233, 174). A similar analysis provided a global spectral fit of $OH^+$ (Müller et al. 2005, J. Mol. Struct., 742, 215). The analysis leading to the detection of $H_2O^+$ occurred in the reverse order – its rotational spectral feature, including a distinctive hyperfine splitting pattern, was first detected in absorption in HIFI spectra (Ossenkopf et al. 2010, A&A 518, L111).  The distinctive hyperfine splitting led to this ion being recognized as a likely carrier, and detailed analysis of its high-resolution infrared spectrum led to the confirmation that this feature arose from $H_2O^+$.

Additional observations have explored the extent of molecular complexity in interstellar and circumstellar environments.  Fullerene molecules such as $C_{60}$ and $C_{70}$ have been prime targets ever since their discovery in laboratory experiments designed to simulate the chemistry of carbon star outflows (Kroto et al. 1985, Nature, 318, 162).  Recent Spitzer Space Telescope observations revealed for the first time the spectroscopic signatures of $C_{60}$ and $C_{70}$ in a variety of astronomical environments.  The molecules were first detected in a hydrogen-poor planetary nebula (Cami et al. 2010, Science, 329, 1180).  The hydrogen-poor conditions were thought to be necessary in light of laboratory measurements on fullerene production (De Vries et al. 1993, Geochim. Cosmochim. Acta, 57, 933; Wang et al. 1995, J. Mater. Res., 10, 1977).  Other Spitzer spectra reveal the presence of $C_{60}$ in the reflection nebulae NGC 7023 and NGC 2023 (Sellgren et al. 2010, ApJ, 722, L54) and in planetary nebulae in our Galaxy and the Small Magellanic Cloud (García-Hernández et al. 2010, ApJ, 724, L39).  The latter work shows that the fullerenes are present in a variety of environments, including hydrogen-rich ones; García-Hernández et al. suggest that the photochemistry of hydrogenated amorphous carbon plays a key role.  These detections would not have been possible without spectroscopic laboratory data (Krätschmer et al. 1990, Nature, 347, 354; Frum et al. 1991, Chem. Phys. Lett., 176, 504; Martin et al. 1993, Phys. Rev., B47, 14607; Nemes et al. 1994, Chem. Phys. Lett., 218, 295; Fabian 1996, Phys. Rev., B53, 13864; Sogoshi et al. 2000, J. Phys. Chem. A, 104, 3733).  These measurements provided the wavelengths and line strengths to confirm the astronomical detections.

In addition to the advances in understanding interstellar and circumstellar chemistry, molecular laboratory studies have led to great improvements in our understanding of the chemistry occurring in the atmospheres of hot-Jupiter exoplanets.  HD 189733b and HD 209458b are two of the most observed hot-Jupiter exoplanets, and recent observations with the NICMOS instrument on the Hubble Space Telescope reveal the presence of molecular features attributed to water, methane, and carbon dioxide in their atmospheres (e.g., Swain et al. 2008, Nature, 452, 329; Swain et al. 2009a, ApJ, 690, L114; Swain et al. 2009b, ApJ, 704, 1616).  Transmission spectra acquired during primary eclipse and dayside emission spectra were the basis for the detections.  Analysis of these spectra provides important constraints on the physical conditions and relevant processes in exoplanet atmospheres.  In order to model the observed spectra, molecular data relevant to high temperatures are required.  Swain et al. (2009b) used data of Barber et al. (2006, MNRAS, 368, 1087) and Zobov et al. (2008, MNRAS, 387, 1093) for the hot water lines.  The results of Nassar & Bernath (2003, J. Quan. Spectrosc. Rad. Transf., 82, 279) and Rothman et al. (2005, J. Quan. Spectrosc. Rad. Transf., 96, 139) were used for $CH_4$ lines.  The compilation of Rothman et al. is known as the HITRAN database.  The data on hot $CO_2$ lines came from Tashkun et al. (2003, J. Quan. Spectrosc. Rad. Transf., 82, 165).

*C.3 Dust and Ices Astrophysics*

During the past few years, observations with the Spitzer Space Telescope have transformed our understanding of dust, polycyclic aromatic hydrocarbons (PAHs), and ices in a wide range of environments, from protoplanetary disks to high-redshift galaxies and even galactic halos. This transformation is a direct result of laboratory studies of dust, PAH, and ice spectroscopy, ice and surface chemistry experiments, and dust coagulation studies.

Spitzer observations of silicate features at 10-30 µm in protoplanetary disks have been crucial to constrain the dust composition, heating history, and coagulation as well as how dust changes radially in the planet-forming zone (e.g., Bouwman et al. 2008, ApJ, 683, 479). These studies used laboratory dust optical constants (e.g., Koike et al. 2003, A&A, 399, 1101). Laboratory dust opacities have been instrumental in quantifying the properties of transition and pre-transition disks among the Spitzer observed disk sample (Calvet et al. 2005, ApJ, 630, L185; Espaillat et al. 2007, ApJ, 670, L135). These disks have large holes or gaps, potentially cleared out by Solar System analogs. Together the dust processing observations and the frequency of these transition objects as well as the sizes of their holes and gaps are among the best constraints available on different (exo)-planet formation models.

Spitzer observed the many mid-infrared spectral features from PAHs toward an unprecedented number of extra-galactic sources across the Universe. These PAH bands, identified and analyzed based on laboratory and computational spectra, have become a standard tool to measure the spectroscopic red-shift, to measure the star formation rate, and to classify the galaxies as either starburst or active galactic nuclei (AGNs) out to redshifts $z > 4$ (e.g., Smith et al. 2007, ApJ, 656, 770; Draine et al. 2007, ApJ, 663, 866; Pope et al. 2008, ApJ, 675, 1171; Riechers et al. 2010, BAAS, 42, 238). Spitzer also discovered a family of new emission features between 15 and 20 µm which, again thanks to the extensive database of PAH spectra produced with support of the Laboratory Astrophysics program, were readily attributed to PAHs. These new features fall in the transition region between nearest neighbor vibrations to full-skeleton modes, giving the first glimpse into the size, geometry, and charge of the largest members of the astronomical PAH family (Bauschlicher et al. 2008, ApJ, 678, 316; Bauschlicher et al, 2009. ApJ, 697, 311; Boersma et al. 2010, A&A, 511, A32). In addition to being critical for understanding mid-IR observations made by JWST, these large PAHs will impact the upcoming far-IR observations made with SOFIA and Herschel.

The importance of different types of interstellar grain surface chemical pathways for the formation of molecules other than $H_2$ has been a topic of debate for decades. The high sensitivity of Spitzer allowed for observations of icy grain mantles toward previously inaccessible astronomical objects such as solar-type protostars, protoplanetary disks, star-less and star-forming clouds, and a range of extra-galactic sources (e.g., Bergin et al. 2005, ApJ, 627, L33; Boogert et al. 2008, ApJ, 678, 985; Oliveira et al. 2009, ApJ, 707, 1269). The ices and ice structures were identified and quantified with laboratory ice spectroscopy (e.g., Hudgins et al. 1993, ApJ, 86, 713; White et al. 2009, ApJS, 180, 182). Together with surface astrochemistry studies on hydrogenation (e.g., Hidaka et al. 2004, ApJ, 614, 1124) these observations have explored the formation efficiencies and pathways of the small molecules commonly associated with the origin of life, especially water, methane, ammonia, and methanol (for a review see Herbst & van Dishoeck 2009, ARAA, 47, 427).

Other laboratory work has established the formation yields and pathways of more complex molecules such as methyl formate and dimethyl ether through UV photolysis and ion bombardment (Bennet et al. 2007, ApJ, 660, 1588; Öberg et al. 2009, A&A, 504, 891). These experiments are being used to interpret the wealth of data on complex organic molecules obtained with Herschel, where enough lines are detected to use these molecules as probes of the physical conditions in hot molecular cores (e.g., Kama et al. 2010, A&A, 521, L39).

*C.4 Plasma Astrophysics*

The most efficient energy sources known in the universe are accretion disks. Those around black holes convert 5-40% of rest mass energy to radiation. Like water circling a drain, inflowing mass must lose angular momentum, presumably by vigorous turbulence in disks, which are essentially inviscid (Shakura & Sunyaev 1973, A&A, 24, 337). The origin of the turbulence is conjectural. Hot disks of electrically conducting plasma inferred from observations of Chandra and XMM (Rappaport et al. 2010, ApJ, 721, 1348) can become turbulent via the linear magnetorotational instability (MRI; Balbus & Hawley 1998, Rev. Mod. Phys., 70, 1). Cool disks, such as the planet-forming disks of protostars inferred from observations with the Hubble Space Telescope and Spitzer (De Marchi et al. 2010, ApJ, 715, 1), may be too poorly ionized for MRI, hence essentially unmagnetized and linearly stable. Nonlinear hydrodynamic instability often occurs in linearly stable flows (e.g., pipe flows) at sufficiently large Reynolds numbers. Although disks have extreme Reynolds numbers, Keplerian rotation enhances their linear hydrodynamic stability, so whether nonmagnetic disks can be turbulent and thereby transport angular momentum effectively is controversial (Richard & Zahn 1999, A&A, 347, 734; Lesur & Longaretti 2005, A&A, 444, 25). Based on a well-controlled and diagnosed laboratory experiment, it was shown (Ji et al. 2006, Nature, 444, 343) that nonmagnetic quasi-Keplerian flows at Reynolds numbers up to millions are essentially steady. Scaled to accretion disks, rates of angular momentum transport lie far below astrophysical requirements. By ruling out purely hydrodynamic turbulence, these results indirectly support MRI as the likely cause of turbulence even in cool disks (Balbus 2009, arXiv:0906.0854).

Laboratory experiments have played an important role in our quest to understand planet formation and planetary interiors. Experiments using different facilities and techniques have unambiguously demonstrated that the transition from non-conducting molecular hydrogen to atomic metallic hydrogen at high pressures is a continuous transition, and not a discontinuous first order phase transition (Nellis 2000, Planet. Space Sci., 48, 671; Celliers et al. 2000, Phys. Rev. Lett., 84, 5564). This suggests that the metallic region of Jupiter's interior extends out to 90% of the radius of the planet, and may explain why the magnetic field of Jupiter is so much stronger than that of the other planets of our solar system. Additionally, connections have recently been established between high-pressure laboratory tests of equation of state (EOS) models of H and He and predictions of the interior structure of Jupiter, from surface to core (Saumon & Guillot 2004, ApJ, 609, 1170; Fortney et al. 2009, Phys. Plasmas, 16, 041003).